\documentclass[twocolumn,showpacs,preprintnumbers,amsmath,amssymb]{revtex4}

\usepackage{graphicx}
\usepackage{dcolumn}
\usepackage{bm}
\usepackage[dvipdf]{hyperref}

\newcommand{\sro}{{$^{88}$}Sr}
\newcommand{\srs}{{$^{86}$}Sr}

\newcommand{\tripl}{$^1$S$_0$-$^3$P$_1$}
\newcommand{\singl}{$^1$S$_0$-$^1$P$_1$}

\newcommand{\meta}{$^3$P$_2$}

\draft
\begin{document}

\title{Cooling of Sr to high phase-space density by laser and sympathetic cooling in isotopic mixtures}

\author{G. Ferrari}
\author{R. E. Drullinger}
\author{N. Poli}
\author{F. Sorrentino}
\author{G. M. Tino}
\email{Guglielmo.Tino@fi.infn.it}

\affiliation{Dipartimento di Fisica and LENS, Istituto Nazionale Fisica Nucleare,
Istituto Nazionale Fisica della Materia,\\
Polo Scientifico-Universit\`a di Firenze, 50019 Sesto Fiorentino, Italy}

\date{\today}

\begin{abstract}
Based on an experimental study of two-body and three-body collisions in ultracold strontium
samples,  a novel optical-sympathetic cooling method in isotopic mixtures is demonstrated. Without
evaporative cooling, a phase-space density of $6\times10^{-2}$ is obtained with a high spatial
density that should allow to overcome the difficulties encountered so far to reach quantum
degeneracy for Sr atoms.

\end{abstract}

 \maketitle

The combination of laser cooling \cite{nobel97} and evaporative  cooling allowed to reach
ultra-low temperatures and to observe Bose-Einstein condensation (BEC) and Fermi degeneracy for
different atoms \cite{Nobel2001bis}. Laser cooling is indeed very effective to reach phase-space
densities within few orders of magnitude of quantum degeneracy. The limits in cooling at high
density are set by the optical depth of the sample, {\it i.e.} reabsorption of scattered light,
and light-assisted atom-atom collisions. Forced evaporative cooling \cite{hess86} represents the
common way to circumvent these limits. However, this procedure does not work for all atoms. In
particular, among the atoms cooled with optical methods, none of the alkali-earth atoms reached
quantum degeneracy so far, except ytterbium which has an alkali-earth-like electronic structure
\cite{takasu03}. A phase-space density of $ \sim 10^{-1}$ was reported for Sr \cite{ido00} but BEC
could not be reached. Sr is of interest for several reasons: In a natural strontium sample,
different isotopes can be investigated with different statistics; $^{88}$Sr (82.6 $\%$), $^{86}$Sr
(9.8 $\%$), and $^{84}$Sr (0.6 $\%$) are bosons with zero nuclear spin; $^{87}$Sr (7$\%$) is a
fermion with nuclear spin $I = 9/2$. A BEC of Sr atoms would allow the study of 0-spin
condensates, fast optical cooling and continuous coherent matter-wave sources, quantum devices
both for frequency standards and inertial sensors. Besides the studies towards quantum degeneracy,
Sr has recently been the subject of active research in various fields spanning from laser cooling
physics \cite{xu03,loftus04}, investigation of ultra-narrow transitions towards future optical
clocks \cite{ferrari03,courtillot03,takamoto03,ido05}, multiple scattering \cite{bidel02}, and
collisional physics \cite{dereviankot03} \cite{santra04} \cite{ciurylo05} \cite{yasuda04}
\cite{mickelson05}.

In this article, we present a  novel scheme for the production of a high phase-space density
sample of atomic strontium. Since collisional processes play a crucial role and relevant
parameters are not well known for this atom, we studied 2-body and 3-body collisions both in pure
samples, and isotopic mixtures. Our results show that usual evaporative cooling methods are
difficult to apply in a single species sample of  \srs\ or \sro. The investigation of an ultracold
\srs-\sro\ mixture revealed a large interspecies elastic cross-section; this allowed us to
demonstrate an optical-sympathetic cooling method conceptually different from what was done so far
on neutral atoms \cite{myatt97,schreck01} and somehow analogous to the one used for trapped ions
\cite{drullinger80}, achieving a phase-space density of 6$\times 10^{-2}$ in conditions suitable
to further cooling of the sample towards BEC.

The experimental setup, based on the apparatus previously described in \cite{ferrari03,poli05},
allows to trap single-isotope samples or isotopic mixtures of \srs\, and \sro\, atoms in an
optical dipole trap. The loading procedure starts with a magneto-optical trap on the \singl\,
resonance transition at 461 nm. During this phase, a decay channel from the excited state results
in the accumulation of atoms in the magnetically trapped metastable \meta\, state \cite{katori00}.
The isotopic mixture is produced in the magnetic trap by stepping the frequency of the trapping
laser to the appropriate values. After loading the magnetic trap, atoms are optically pumped back
into the electronic ground state and a second cooling stage is applied based on a magneto-optical
trap operating on the narrow \tripl\, transition at $689$ nm. At this point we typically obtain
$10^7$ \sro\, atoms at 2 $\mu$K, and $10^6$ \srs\, atoms at 1 $\mu$K. The isotopic mixture is then
loaded into an optical dipole trap by superposing a focused 922 nm laser on the red
magneto-optical trap, resulting in a 90 $\mu$K deep potential with radial and axial frequencies of
2 kHz and 26 Hz, respectively. To optimize the transfer efficiency from the red magneto-optical
trap into the optical dipole trap, the optical dipole beam waist is axially displaced from the
magneto-optical trap by 500 $\mu$m (corresponding to 2/3 of the dipole beam Rayleigh range and two
times the red magneto-optical trap radii) in order to increase the magneto-optical trap-dipole
trap spatial overlap \cite{kuppens00}. The measured 6 s lifetime in the optical trap is consistent
with the residual background gas pressure in our apparatus. The off-resonant photon scattering
induces a heating of 0.4 $\mu$K/s. The polarization of the dipole beam is set to nearly fulfill
the requirements for trapping at the "{\it magic wavelength}" that cancels the differential Stark
shift for the \tripl\, transition \cite{ido03}. The atom number and temperature are measured
independently on the two isotopes by absorption imaging with a resonant probe beam, and the
contribution of the non-resonant isotope is taken into account. The spatial density is inferred
from the measured temperature and the trapping potential.

Two-body and three-body collisions are studied by loading single species or the isotopic mixture
into the optical dipole trap. For the investigation of 2-body elastic collisions, the gas is
initially put out of equilibrium by a sudden perturbation of the optical potential (pure samples),
or by an isotope-selective laser cooling pulse (mixture). We relate the measured thermalization
time at the given density and temperature to the elastic cross-section $\sigma$ following and
generalizing the approach described in \cite{arndt97}. The resulting values are
$\sigma_{88-88}=(3\pm 1) \times 10^{-17}$ m$^2$, $\sigma_{86-86}=(1.3\pm 0.5) \times 10^{-14}$
m$^2$, and $\sigma_{86-88}=(4\pm 1) \times 10^{-16}$ m$^2$. In the single isotope case, the
results are consistent with the corresponding scattering lengths measured by photoassociation
spectroscopy \cite{yasuda04,mickelson05}. No previous reference data is available for the
two-isotope collisions.

\begin{figure}[t]
\vspace{0mm} \begin{center}
\includegraphics[width=0.4\textwidth,angle=90]{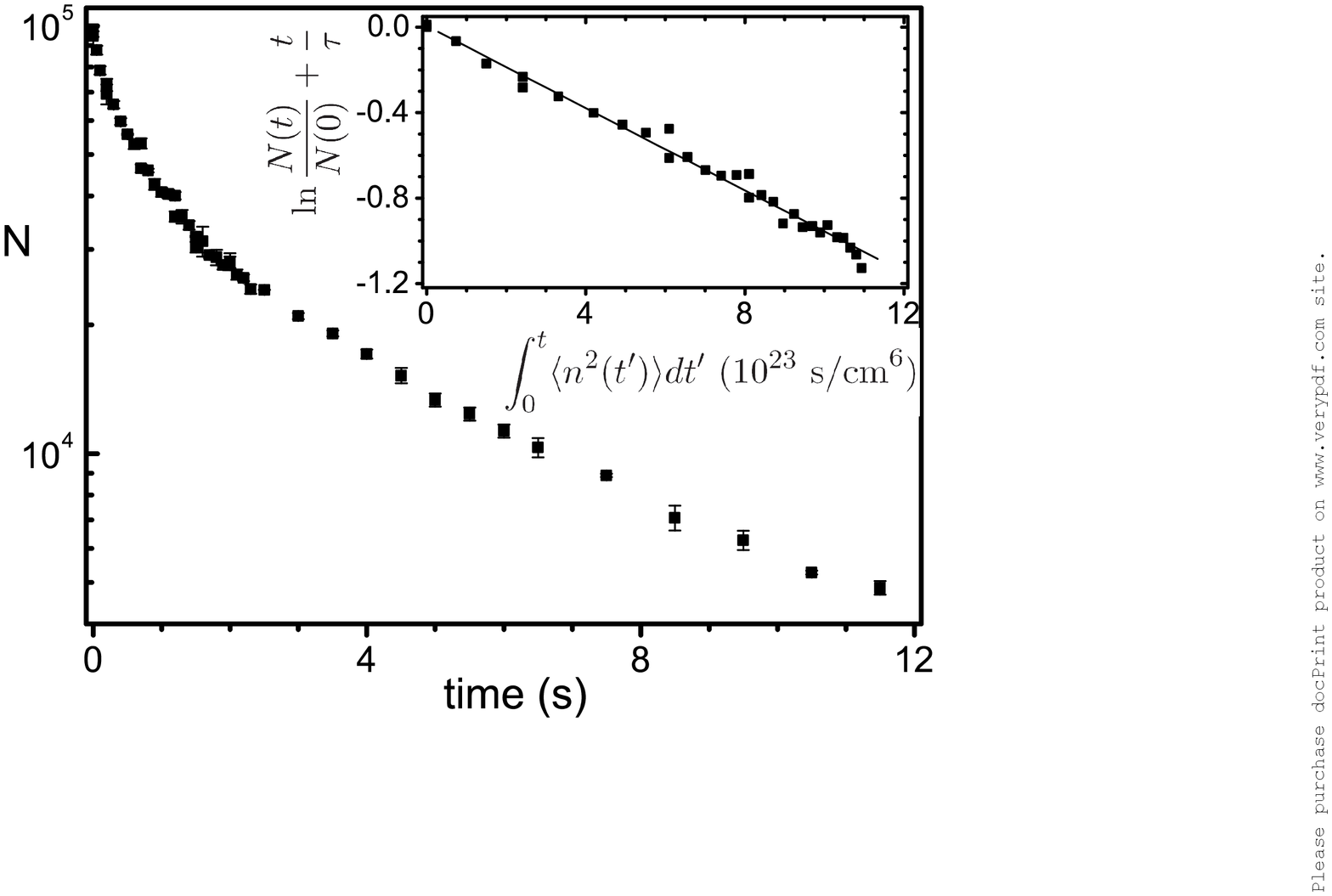}
\vspace{-14mm} \caption{\label{GraphDecay86-86} Non-exponential decay of the number
of \srs, trapped in the optical dipole trap due to 3-body recombination. Inset: the 3-body recombination rate constant is given by the slope of the natural log of the atom number with respect to $\int^t_0\langle n^2(t')\rangle dt'$ where $\langle n^2(t')\rangle=\frac{1}{N(t')}\int_Vn^3(\vec{r},t')d^3r$ (see the text). The term $t/\tau$ accounts for losses due to background collisions.}
\end{center}
\end{figure}

We studied the 3-body recombination rate $K_3$ by loading a single isotope (either \srs\ or \sro)
into the dipole trap and observing the evolution of the atom number and temperature. With \sro\ we
found no experimental evidence for non-exponential decay and, given the initial atom density at
the trap center of $3\times 10^{13}$ cm$^{-3}$, we set an upper bound on $K_3^{88}$ at $10^{-27}$
cm$^6$/s. To determine $K_3^{86}$ we followed the analysis described in \cite{burt97}: the loss
rate due to 3-body recombination is modeled by the rate equation
\begin{equation}\label{rateEq}
\frac{dN}{dt}=-K_3\int_V n^3(\vec{r},t)d^3r
\end{equation}
or equivalently
\begin{equation}\label{rateEqIntegrate}
\ln\frac{N(t)}{N(0)}=-K_3\int_0^t dt'\int_V\frac{n^3(\vec{r},t')}{N(t')}d^3r,
\end{equation}
where $n$ is the density, and $N(t)$ is the atom number at time $t$. Figure \ref{GraphDecay86-86}
shows the data and the fit using equation \ref{rateEqIntegrate} from which we obtain the rate
constant $K_3^{86}= (1.0\pm0.5)\times 10^{-24}$ cm$^6$/s. This recombination rate is more than
three orders of magnitude larger than the typical values for ultracold alkali vapors
\cite{moerdijk96}.

The first consequence of these results is that evaporative cooling on pure samples of either \srs\
or \sro\ cannot be efficient. \srs\ presents an extremely large elastic cross-section, a good
point to establish a fast thermalization, but the 3-body recombination rate introduces a loss
channel that is fatal with the typical geometries accessible through optical dipole trapping. An
optical trap with a much larger trapping volume would partially suppress this loss channel
\cite{weber02}.  \sro\ instead turns out to be stable against 3-body decay, but the small elastic
cross-section results in a long thermalization time compared to typical trap lifetime. Indeed, in
our experiment evaporative cooling of \sro\ by reducing the dipole beam intensity allowed to
increase the phase-space density to a maximum of $2\times 10^{-1}$, limited by the reduction of
the thermalization rate during the evaporation and by the 6 s background limited lifetime.

These results led to a new all-optical sympathetic cooling scheme. The basic idea of the new
cooling method is to use a mixture of isotopes trapped in the dipole trap: one isotope is laser
cooled and by elastic collisions cools the other isotope which is the one of interest. The
advantage, compared to the schemes used so far for neutral atoms, is that the limitations of laser
cooling of spatially dense and optically thick samples are drastically reduced. In this work,
continuous laser cooling of \srs\ leads to heat dissipation in a \sro\ sample by sympathetic
cooling; the small \srs\ optical depth does not limit the achievable minimum temperature.
Sympathetic cooling with neutral atoms normally requires a thermal bath with a large heat capacity
with respect to that of the sample to be cooled. This is due to the fact that when the thermal
bath is cooled by evaporative cooling, each lost atom carries an energy of the order of few times
the temperature of the sample. In the case of optical-sympathetic cooling, each laser-cooled atom
can subtract an energy of the order of the optical recoil in a time corresponding to a few
lifetimes of the excited state (say 4, when operating at 1/4 $I_{sat}$), without being lost.
Considering the \tripl\, transition, the power subtracted per atom is of the order of 6 mK/s; if
the inter-isotope thermalization is fast enough, each \srs\, atom can cool about 10$^{3}$ \sro\,
atoms down to $\mu$K temperatures in one second. In the \srs-\sro\ isotope mixture the relatively
large inter-species cross-section results in thermalization times typically of the order of few
milliseconds. This thermalization is fast even on the time scale of laser cooling on the
intercombination \tripl\ transition. On the other hand, the 164 MHz \srs-\sro\ isotopic shift and
the natural linewidth of 7.6 kHz  for the \tripl\ transition, makes laser cooling on one isotope
insensitive to the presence of the other.

\begin{figure}[t]
\vspace{0mm}

\hspace{-0mm}
\includegraphics[width=0.34\textwidth,angle=90]{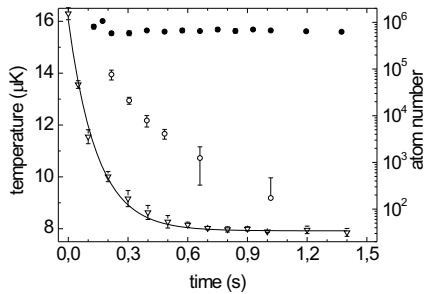}

\vspace{-3mm} \caption{\label{OptSympatheticDynamics} Dynamics of an optically trapped \sro\,
cloud sympathetically cooled with laser cooled \srs. Filled circles $\bullet$: \sro\ atom number.
Open circles $\circ$: \srs\ atom number. The number of \sro\, atoms remains constant during the
process, while \srs\, decays exponentially with a 80 ms time constant. Under optimized conditions,
the temperature (triangles $\triangledown$) decreases with a 150 ms time constant and the mixture
is always at thermal equilibrium. }

\end{figure}

\begin{figure}[t] \vspace{-0mm} \begin{center}
\includegraphics[width=0.34\textwidth,angle=90]{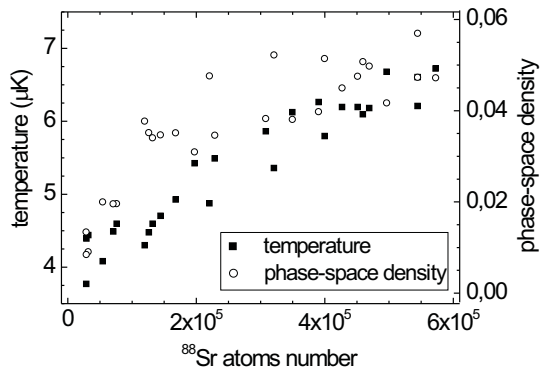} \vspace{-13mm} \caption{\label{LimitTempVsNum}
Temperature and phase-space density of the \sro\, cloud sympathetically cooled with laser cooled
\srs, as a function of the \sro\, atom number.}
\end{center}
\end{figure}

We implemented the optical-sympathetic cooling scheme by extending the temporal overlap between
the optical dipole trap and the \srs\ red magneto-optical trap after switching off the \sro\
magneto-optical trap.  During this phase the parameters that optimize the \sro\, cooling are
substantially the same as those typically employed in the second cooling stage (intensity 30-100
$I_{sat}$ with $I_{sat}=\,3\mu$W/cm$^2$, -200 kHz frequency detuning, magnetic field gradient 0.6
mT/cm). The horizontal displacement between the red magneto-optical trap and the  dipole trap
optimizes the spatial overlap and provides a continuous flux of \srs\ atoms from the
magneto-optical trap to the optical dipole trap replacing the atoms that are lost during the
optical-sympathetic cooling due to light assisted collisions. Figure \ref{OptSympatheticDynamics}
reports the dynamics of optical-sympathetic cooling, starting from an initial temperature of 15-20
$\mu$K, limited by density dependent heating \cite{katori99,poli05}. It can be observed that the
cooling process does not induce losses on \sro\ while the number of \srs\, atoms exponentially
decays with a 80 ms lifetime. About 100 ms after switching off the \sro\, red magneto-optical
trap, we observe that the mixture attains thermal equilibrium. From this point on the equilibrium
is maintained, which is an essential condition for the implementation of this cooling scheme.
Under optimized conditions (overall optical intensity 100 $I_{sat}$) the temperature decays from
the initial value with a 150 ms time constant. The minimum attainable temperature depends both on
the intensity of the \srs\, cooling beam, and the \sro\, density.
\begin{figure}[t]
\vspace{0mm} \begin{center}
\includegraphics[width=0.34\textwidth,angle=90]{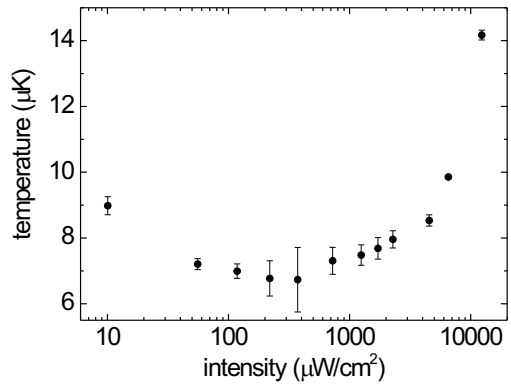}
\vspace{-12mm} \caption{\label{TempVsFinalInt} Asymptotic temperature of the \sro\, cloud for different intensities of the \srs\, laser cooling beam on the \tripl\, transition ($I_{sat}$ = 3 $\mu$W/cm$^2$). The \sro\, atom number is $6\times10^5$.}
\end{center}
\end{figure}
Figure \ref{TempVsFinalInt} reports the asymptotic temperature as a function of the overall
cooling light intensity incident on the sample, for a fixed amount of trapped \sro. At high
intensity, the \srs\, lifetime increases due to the reduced spatial density and the final
temperature is proportional to the intensity as expected from the usual Doppler theory. By
lowering the intensity, the final temperature and the \srs\, lifetime decrease  to the point where
\srs\, is lost too fast with respect to the cooling dynamics of \sro. The optimum values are
obtained when the overall cooling light incident on the atoms has an intensity $\sim 30 \,
I_{sat}$. For $6\times 10^5$ \sro\ atoms trapped in the dipole trap, the  final temperature is
$6.7\mu$K at a peak density of $1.3\times10^{14}$ cm$^{3}$; the corresponding phase-space density
is $5\times10^{-2}$. This value is only a factor of two lower with respect to what was previously
obtained without forced evaporation \cite{ido00}, but it exhibits more favorable conditions for
starting evaporative cooling, considering the higher spatial density (more than one order of
magnitude higher) and the larger number of trapped atoms. By keeping the cooling parameters on
\srs\, fixed at the optimum value and by varying the number of trapped \sro, we determined the
dependence of the final temperature with respect to the \sro\, density. Figure
\ref{LimitTempVsNum} shows the dependence of \sro\, temperature from  the number of atoms in the
trap. We determine the density-dependent heating $dT/dn\simeq 2\, \mu$K/($10^{14}$ cm$^{-3}$),
which is 20 times lower than the equivalent value for pure laser cooled \sro\ \cite{katori99}.
This strong reduction is a direct consequence of the optical distinction of the two isotopes on
the \tripl\, transition. The limit on the \srs\ temperature of 4 $\mu$K for zero \sro\ density can
be attributed to the laser cooling dynamics in the tightly confining potential of the dipole trap.
Considering the power laws involved, reducing the optical dipole potential in the final stage of
the optical-sympathetic cooling should lead to further cooling of the sample and considerably
increase the \sro\ phase-space density.

In conclusion, we investigated the collisional properties of ultracold Sr atoms of interest for
the production of a Bose-Einstein condensate. On pure \srs\ we find a very large 3-body
recombination rate which sets stringent limits on the evaporation efficiency on this isotope. For
\sro\ we find that the small elastic cross-section results in a long thermalization time compared
to typical trap lifetime.  This can explain why BEC of Sr was not achieved so far. On the other
hand, for the \srs-\sro\ mixture we measure a rather large inter-species elastic cross section. We
exploit the fast \srs-\sro\ thermalization  to demonstrate an optical-sympathetic cooling scheme
in which heat dissipation for one isotope is obtained by laser cooling; the sympathetically cooled
component is more than one order of magnitude more abundant than the laser cooled one, reducing
substantially the limitations associated to laser cooling at high spatial density. We cool up to
$7\times10^{5}$ \sro\ atoms at a peak spatial density of $1.4\times10^{14}$ cm$^{-3}$ and $6\times
10^{-2}$ phase space density.  The combination of this cooling scheme to a final evaporative
cooling stage on pure \sro \, should allow to increase considerably the phase-space density
opening the possibility to investigate  quantum degenerate gases of Sr. An interesting possibility
is also the development of new atomic sensors based on ultracold Sr atoms exploting the weak
interatomic interaction and their small sensitivity to magnetic and electric fields.

We acknowledge fruitful discussions with M. Prevedelli and A. Simoni. We thank R. Ballerini, M. De
Pas, M. Giuntini, and A. Hajeb for technical assistance. This work was supported by Agenzia
Spaziale Italiana, Ente Cassa di Risparmio di Firenze, MIUR-COFIN 2002, Istituto Nazionale Fisica
della Materia, Istituto Nazionale di Fisica Nucleare, and LENS.


\begin{thebibliography}{29}
\expandafter\ifx\csname natexlab\endcsname\relax\def\natexlab#1{#1}\fi \expandafter\ifx\csname
bibnamefont\endcsname\relax
  \def\bibnamefont#1{#1}\fi
\expandafter\ifx\csname bibfnamefont\endcsname\relax
  \def\bibfnamefont#1{#1}\fi
\expandafter\ifx\csname citenamefont\endcsname\relax
  \def\citenamefont#1{#1}\fi
\expandafter\ifx\csname url\endcsname\relax
  \def\url#1{\texttt{#1}}\fi
\expandafter\ifx\csname urlprefix\endcsname\relax\def\urlprefix{URL }\fi
\providecommand{\bibinfo}[2]{#2} \providecommand{\eprint}[2][]{\url{#2}}

\bibitem[{\citenamefont{Chu et~al.}(1998)\citenamefont{Chu, Cohen-Tannoudji,
  and Phillips}}]{nobel97}
\bibinfo{author}{\bibfnamefont{S.}~\bibnamefont{Chu}},
  \bibinfo{author}{\bibfnamefont{C.}~\bibnamefont{Cohen-Tannoudji}},
  \bibnamefont{and} \bibinfo{author}{\bibfnamefont{W.~D.}
  \bibnamefont{Phillips}}, \bibinfo{journal}{{\it Nobel Lectures in Physics
  1997}, Rev. Mod. Phys.} \textbf{\bibinfo{volume}{70}}, \bibinfo{pages}{685}
  (\bibinfo{year}{1998}).

\bibitem[{Nob()}]{Nobel2001bis}
\emph{\bibinfo{title}{{\rm E. A. Cornell and C. E. Wieman, {\it Nobel Lectures
  in Physics 2001}, Rev. Mod. Phys. {\bf 74}, 875, 1131 (2002); W. Ketterle,
  {\it ibid.}, 1131 (2002)}}}.

\bibitem[{\citenamefont{Hess}(1986)}]{hess86}
\bibinfo{author}{\bibfnamefont{H.~F.} \bibnamefont{Hess}},
  \bibinfo{journal}{Phys. Rev. B} \textbf{\bibinfo{volume}{34}},
  \bibinfo{pages}{(R)3476} (\bibinfo{year}{1986}).

\bibitem[{\citenamefont{Takasu et~al.}(2003)\citenamefont{Takasu, Maki, Komori,
  Takano, Honda, Kumakura, Yabuzaki, and Takahashi}}]{takasu03}
\bibinfo{author}{\bibfnamefont{Y.}~\bibnamefont{Takasu}},
  \bibinfo{author}{\bibfnamefont{K.}~\bibnamefont{Maki}},
  \bibinfo{author}{\bibfnamefont{K.}~\bibnamefont{Komori}},
  \bibinfo{author}{\bibfnamefont{T.}~\bibnamefont{Takano}},
  \bibinfo{author}{\bibfnamefont{K.}~\bibnamefont{Honda}},
  \bibinfo{author}{\bibfnamefont{M.}~\bibnamefont{Kumakura}},
  \bibinfo{author}{\bibfnamefont{T.}~\bibnamefont{Yabuzaki}}, \bibnamefont{and}
  \bibinfo{author}{\bibfnamefont{Y.}~\bibnamefont{Takahashi}},
  \bibinfo{journal}{Phys. Rev. Lett.} \textbf{\bibinfo{volume}{91}},
  \bibinfo{pages}{040404} (\bibinfo{year}{2003}).

\bibitem[{\citenamefont{Ido et~al.}(2000)\citenamefont{Ido, Isoya, and
  Katori}}]{ido00}
\bibinfo{author}{\bibfnamefont{T.}~\bibnamefont{Ido}},
  \bibinfo{author}{\bibfnamefont{Y.}~\bibnamefont{Isoya}}, \bibnamefont{and}
  \bibinfo{author}{\bibfnamefont{H.}~\bibnamefont{Katori}},
  \bibinfo{journal}{Phys. Rev. A} \textbf{\bibinfo{volume}{61}},
  \bibinfo{pages}{061403(R)} (\bibinfo{year}{2000}).

\bibitem[{\citenamefont{Xu et~al.}(2003)\citenamefont{Xu, Loftus, Dunn, Greene,
  Hall, Gallagher, and Ye}}]{xu03}
\bibinfo{author}{\bibfnamefont{X.}~\bibnamefont{Xu}},
  \bibinfo{author}{\bibfnamefont{T.~H.} \bibnamefont{Loftus}},
  \bibinfo{author}{\bibfnamefont{J.~W.} \bibnamefont{Dunn}},
  \bibinfo{author}{\bibfnamefont{C.~H.} \bibnamefont{Greene}},
  \bibinfo{author}{\bibfnamefont{J.~L.} \bibnamefont{Hall}},
  \bibinfo{author}{\bibfnamefont{A.}~\bibnamefont{Gallagher}},
  \bibnamefont{and} \bibinfo{author}{\bibfnamefont{J.}~\bibnamefont{Ye}},
  \bibinfo{journal}{Phys. Rev. Lett.} \textbf{\bibinfo{volume}{90}},
  \bibinfo{pages}{193002} (\bibinfo{year}{2003}).

\bibitem[{\citenamefont{Loftus et~al.}(2004)\citenamefont{Loftus, Ido, Ludlow,
  Boyd, and Ye}}]{loftus04}
\bibinfo{author}{\bibfnamefont{T.~H.} \bibnamefont{Loftus}},
  \bibinfo{author}{\bibfnamefont{T.}~\bibnamefont{Ido}},
  \bibinfo{author}{\bibfnamefont{A.~D.} \bibnamefont{Ludlow}},
  \bibinfo{author}{\bibfnamefont{M.~M.} \bibnamefont{Boyd}}, \bibnamefont{and}
  \bibinfo{author}{\bibfnamefont{J.}~\bibnamefont{Ye}}, \bibinfo{journal}{Phys.
  Rev. Lett.} \textbf{\bibinfo{volume}{93}}, \bibinfo{pages}{073003}
  (\bibinfo{year}{2004}).

\bibitem[{\citenamefont{Ferrari et~al.}(2003)\citenamefont{Ferrari, Cancio,
  Drullinger, Giusfredi, Poli, Prevedelli, Toninelli, and Tino}}]{ferrari03}
\bibinfo{author}{\bibfnamefont{G.}~\bibnamefont{Ferrari}},
  \bibinfo{author}{\bibfnamefont{P.}~\bibnamefont{Cancio}},
  \bibinfo{author}{\bibfnamefont{R.~E.} \bibnamefont{Drullinger}},
  \bibinfo{author}{\bibfnamefont{G.}~\bibnamefont{Giusfredi}},
  \bibinfo{author}{\bibfnamefont{N.}~\bibnamefont{Poli}},
  \bibinfo{author}{\bibfnamefont{M.}~\bibnamefont{Prevedelli}},
  \bibinfo{author}{\bibfnamefont{C.}~\bibnamefont{Toninelli}},
  \bibnamefont{and} \bibinfo{author}{\bibfnamefont{G.~M.} \bibnamefont{Tino}},
  \bibinfo{journal}{Phys. Rev. Lett.} \textbf{\bibinfo{volume}{91}},
  \bibinfo{pages}{243002} (\bibinfo{year}{2003}).

\bibitem[{\citenamefont{Courtillot et~al.}(2003)\citenamefont{Courtillot,
  Quessada, Kovacich, Brusch, Kolker, Zondy, Rovera, and
  Lemonde}}]{courtillot03}
\bibinfo{author}{\bibfnamefont{I.}~\bibnamefont{Courtillot}},
  \bibinfo{author}{\bibfnamefont{A.}~\bibnamefont{Quessada}},
  \bibinfo{author}{\bibfnamefont{R.~P.} \bibnamefont{Kovacich}},
  \bibinfo{author}{\bibfnamefont{A.}~\bibnamefont{Brusch}},
  \bibinfo{author}{\bibfnamefont{D.}~\bibnamefont{Kolker}},
  \bibinfo{author}{\bibfnamefont{J.-J.} \bibnamefont{Zondy}},
  \bibinfo{author}{\bibfnamefont{G.~D.} \bibnamefont{Rovera}},
  \bibnamefont{and} \bibinfo{author}{\bibfnamefont{P.}~\bibnamefont{Lemonde}},
  \bibinfo{journal}{Phys. Rev. A} \textbf{\bibinfo{volume}{68}},
  \bibinfo{pages}{030501} (\bibinfo{year}{2003}).

\bibitem[{\citenamefont{Takamoto and Katori}(2003)}]{takamoto03}
\bibinfo{author}{\bibfnamefont{M.}~\bibnamefont{Takamoto}} \bibnamefont{and}
  \bibinfo{author}{\bibfnamefont{H.}~\bibnamefont{Katori}},
  \bibinfo{journal}{Phys. Rev. Lett.} \textbf{\bibinfo{volume}{91}},
  \bibinfo{pages}{223001} (\bibinfo{year}{2003}).

\bibitem[{\citenamefont{Ido et~al.}(2005)\citenamefont{Ido, Loftus, Boyd,
  Ludlow, Holman, and Ye}}]{ido05}
\bibinfo{author}{\bibfnamefont{T.}~\bibnamefont{Ido}},
  \bibinfo{author}{\bibfnamefont{T.~H.} \bibnamefont{Loftus}},
  \bibinfo{author}{\bibfnamefont{M.~M.} \bibnamefont{Boyd}},
  \bibinfo{author}{\bibfnamefont{A.~D.} \bibnamefont{Ludlow}},
  \bibinfo{author}{\bibfnamefont{K.~W.} \bibnamefont{Holman}},
  \bibnamefont{and} \bibinfo{author}{\bibfnamefont{J.}~\bibnamefont{Ye}},
  \bibinfo{journal}{Phys. Rev. Lett.} \textbf{\bibinfo{volume}{94}},
  \bibinfo{pages}{153001} (\bibinfo{year}{2005}).

\bibitem[{\citenamefont{Bidel et~al.}(2002)\citenamefont{Bidel, Klappauf,
  Bernard, Delande, Labeyrie, Miniatura, Wilkowski, and Kaiser}}]{bidel02}
\bibinfo{author}{\bibfnamefont{Y.}~\bibnamefont{Bidel}},
  \bibinfo{author}{\bibfnamefont{B.}~\bibnamefont{Klappauf}},
  \bibinfo{author}{\bibfnamefont{J.~C.} \bibnamefont{Bernard}},
  \bibinfo{author}{\bibfnamefont{D.}~\bibnamefont{Delande}},
  \bibinfo{author}{\bibfnamefont{G.}~\bibnamefont{Labeyrie}},
  \bibinfo{author}{\bibfnamefont{C.}~\bibnamefont{Miniatura}},
  \bibinfo{author}{\bibfnamefont{D.}~\bibnamefont{Wilkowski}},
  \bibnamefont{and} \bibinfo{author}{\bibfnamefont{R.}~\bibnamefont{Kaiser}},
  \bibinfo{journal}{Phys. Rev. Lett.} \textbf{\bibinfo{volume}{88}},
  \bibinfo{pages}{203902} (\bibinfo{year}{2002}).

\bibitem[{\citenamefont{Derevianko et~al.}(2003)\citenamefont{Derevianko,
  Porsev, Kotochigova, Tiesinga, and Julienne}}]{dereviankot03}
\bibinfo{author}{\bibfnamefont{A.}~\bibnamefont{Derevianko}},
  \bibinfo{author}{\bibfnamefont{S.~G.} \bibnamefont{Porsev}},
  \bibinfo{author}{\bibfnamefont{S.}~\bibnamefont{Kotochigova}},
  \bibinfo{author}{\bibfnamefont{E.}~\bibnamefont{Tiesinga}}, \bibnamefont{and}
  \bibinfo{author}{\bibfnamefont{P.~S.} \bibnamefont{Julienne}},
  \bibinfo{journal}{Phys. Rev. Lett.} \textbf{\bibinfo{volume}{87}},
  \bibinfo{pages}{023002} (\bibinfo{year}{2003}).

\bibitem[{\citenamefont{Santra et~al.}(2004, and references
  therein)\citenamefont{Santra, Christ, and Greene}}]{santra04}
\bibinfo{author}{\bibfnamefont{R.}~\bibnamefont{Santra}},
  \bibinfo{author}{\bibfnamefont{K.~V.} \bibnamefont{Christ}},
  \bibnamefont{and} \bibinfo{author}{\bibfnamefont{C.~H.}
  \bibnamefont{Greene}}, \bibinfo{journal}{Phys. Rev. A}
  \textbf{\bibinfo{volume}{69}}, \bibinfo{pages}{042510} (\bibinfo{year}{2004}),
  and references therein.

\bibitem[{\citenamefont{Ciury{\l}o et~al.}(2005)\citenamefont{Ciury{\l}o,
  Tiesinga, and Julienne}}]{ciurylo05}
\bibinfo{author}{\bibfnamefont{R.}~\bibnamefont{Ciury{\l}o}},
  \bibinfo{author}{\bibfnamefont{E.}~\bibnamefont{Tiesinga}}, \bibnamefont{and}
  \bibinfo{author}{\bibfnamefont{P.~S.} \bibnamefont{Julienne}},
  \bibinfo{journal}{Phys. Rev. A} \textbf{\bibinfo{volume}{71}},
  \bibinfo{pages}{030701} (\bibinfo{year}{2005}).

\bibitem[{\citenamefont{Yasuda et~al.}(2005)\citenamefont{Yasuda, Kishimoto,
  Takamoto, and Katori}}]{yasuda04}
\bibinfo{author}{\bibfnamefont{M.}~\bibnamefont{Yasuda}},
  \bibinfo{author}{\bibfnamefont{T.}~\bibnamefont{Kishimoto}},
  \bibinfo{author}{\bibfnamefont{M.}~\bibnamefont{Takamoto}}, \bibnamefont{and}
  \bibinfo{author}{\bibfnamefont{H.}~\bibnamefont{Katori}},
  \bibinfo{journal}{arXiv:physics/0501053}  (\bibinfo{year}{2005}).

\bibitem[{\citenamefont{Mickelson et~al.}(2005)\citenamefont{Mickelson,
  Martinez, Saenz, Nagel, Chen, Killian, Pellegrini, and
  C\^{o}t\'{e}}}]{mickelson05}
\bibinfo{author}{\bibfnamefont{P.~G.} \bibnamefont{Mickelson}},
  \bibinfo{author}{\bibfnamefont{Y.~N.} \bibnamefont{Martinez}},
  \bibinfo{author}{\bibfnamefont{A.~D.} \bibnamefont{Saenz}},
  \bibinfo{author}{\bibfnamefont{S.~B.} \bibnamefont{Nagel}},
  \bibinfo{author}{\bibfnamefont{Y.~C.} \bibnamefont{Chen}},
  \bibinfo{author}{\bibfnamefont{T.~C.} \bibnamefont{Killian}},
  \bibinfo{author}{\bibfnamefont{P.}~\bibnamefont{Pellegrini}},
  \bibnamefont{and}
  \bibinfo{author}{\bibfnamefont{R.}~\bibnamefont{C\^{o}t\'{e}}},
  \bibinfo{journal}{arXiv:physics/0507112}  (\bibinfo{year}{2005}).

\bibitem[{\citenamefont{Myatt et~al.}(1997)\citenamefont{Myatt, Burt, Ghrist,
  Cornell, and Wieman}}]{myatt97}
\bibinfo{author}{\bibfnamefont{C.~J.} \bibnamefont{Myatt}},
  \bibinfo{author}{\bibfnamefont{E.~A.} \bibnamefont{Burt}},
  \bibinfo{author}{\bibfnamefont{R.~W.} \bibnamefont{Ghrist}},
  \bibinfo{author}{\bibfnamefont{E.~A.} \bibnamefont{Cornell}},
  \bibnamefont{and} \bibinfo{author}{\bibfnamefont{C.~E.}
  \bibnamefont{Wieman}}, \bibinfo{journal}{Phys. Rev. Lett.}
  \textbf{\bibinfo{volume}{78}}, \bibinfo{pages}{586} (\bibinfo{year}{1997}).

\bibitem[{\citenamefont{Schreck et~al.}(2001)\citenamefont{Schreck, Ferrari,
  Corwin, Cubizolles, Khaykovich, Mewes, and Salomon}}]{schreck01}
\bibinfo{author}{\bibfnamefont{F.}~\bibnamefont{Schreck}},
  \bibinfo{author}{\bibfnamefont{G.}~\bibnamefont{Ferrari}},
  \bibinfo{author}{\bibfnamefont{K.~L.} \bibnamefont{Corwin}},
  \bibinfo{author}{\bibfnamefont{J.}~\bibnamefont{Cubizolles}},
  \bibinfo{author}{\bibfnamefont{L.}~\bibnamefont{Khaykovich}},
  \bibinfo{author}{\bibfnamefont{M.-O.} \bibnamefont{Mewes}}, \bibnamefont{and}
  \bibinfo{author}{\bibfnamefont{C.}~\bibnamefont{Salomon}},
  \bibinfo{journal}{Phys. Rev. A} \textbf{\bibinfo{volume}{64}},
  \bibinfo{pages}{011402(R)} (\bibinfo{year}{2001}).

\bibitem[{\citenamefont{Drullinger et~al.}(1980)\citenamefont{Drullinger,
  Wineland, and Bergquist}}]{drullinger80}
\bibinfo{author}{\bibfnamefont{R.~E.} \bibnamefont{Drullinger}},
  \bibinfo{author}{\bibfnamefont{D.~J.} \bibnamefont{Wineland}},
  \bibnamefont{and} \bibinfo{author}{\bibfnamefont{J.~C.}
  \bibnamefont{Bergquist}}, \bibinfo{journal}{Appl. Phys.}
  \textbf{\bibinfo{volume}{22}}, \bibinfo{pages}{365} (\bibinfo{year}{1980}).

\bibitem[{\citenamefont{Poli et~al.}(2005)\citenamefont{Poli, Drullinger,
  Ferrari, L\'eonard, Sorrentino, and Tino}}]{poli05}
\bibinfo{author}{\bibfnamefont{N.}~\bibnamefont{Poli}},
  \bibinfo{author}{\bibfnamefont{R.~E.} \bibnamefont{Drullinger}},
  \bibinfo{author}{\bibfnamefont{G.}~\bibnamefont{Ferrari}},
  \bibinfo{author}{\bibfnamefont{J.}~\bibnamefont{L\'eonard}},
  \bibinfo{author}{\bibfnamefont{F.}~\bibnamefont{Sorrentino}},
  \bibnamefont{and} \bibinfo{author}{\bibfnamefont{G.~M.} \bibnamefont{Tino}},
  \bibinfo{journal}{Phys. Rev. A} \textbf{\bibinfo{volume}{71}},
  \bibinfo{pages}{061403(R)} (\bibinfo{year}{2005}).

\bibitem[{\citenamefont{Katori et~al.}(2001)\citenamefont{Katori, Ido, Isoda,
  and Kuwata-Gonokami}}]{katori00}
\bibinfo{author}{\bibfnamefont{H.}~\bibnamefont{Katori}},
  \bibinfo{author}{\bibfnamefont{T.}~\bibnamefont{Ido}},
  \bibinfo{author}{\bibfnamefont{Y.}~\bibnamefont{Isoda}}, \bibnamefont{and}
  \bibinfo{author}{\bibfnamefont{M.}~\bibnamefont{Kuwata-Gonokami}}, in
  \emph{\bibinfo{booktitle}{Atomic Physics 17}}, edited by
  \bibinfo{editor}{\bibfnamefont{E.}~\bibnamefont{Arimondo}},
  \bibinfo{editor}{\bibfnamefont{P.}~\bibnamefont{de~Natale}},
  \bibnamefont{and} \bibinfo{editor}{\bibfnamefont{M.}~\bibnamefont{Inguscio}}
  (\bibinfo{publisher}{AIP},
  \bibinfo{year}{2001}), p. \bibinfo{pages}{382}.

\bibitem[{\citenamefont{Kuppens et~al.}(2000)\citenamefont{Kuppens, Corwin,
  Miller, Chupp, and Wieman}}]{kuppens00}
\bibinfo{author}{\bibfnamefont{S.~J.~M.} \bibnamefont{Kuppens}},
  \bibinfo{author}{\bibfnamefont{K.~L.} \bibnamefont{Corwin}},
  \bibinfo{author}{\bibfnamefont{K.~W.} \bibnamefont{Miller}},
  \bibinfo{author}{\bibfnamefont{T.~E.} \bibnamefont{Chupp}}, \bibnamefont{and}
  \bibinfo{author}{\bibfnamefont{C.~E.} \bibnamefont{Wieman}},
  \bibinfo{journal}{Phys. Rev. A} \textbf{\bibinfo{volume}{62}},
  \bibinfo{pages}{013406} (\bibinfo{year}{2000}).

\bibitem[{\citenamefont{Ido and Katori}(2003)}]{ido03}
\bibinfo{author}{\bibfnamefont{T.}~\bibnamefont{Ido}} \bibnamefont{and}
  \bibinfo{author}{\bibfnamefont{H.}~\bibnamefont{Katori}},
  \bibinfo{journal}{Phys. Rev. Lett.} \textbf{\bibinfo{volume}{91}},
  \bibinfo{pages}{053001} (\bibinfo{year}{2003}).

\bibitem[{\citenamefont{Arndt et~al.}(1997)\citenamefont{Arndt, Dahan,
  Gu\'ery-Odelin, Reynolds, and Dalibard}}]{arndt97}
\bibinfo{author}{\bibfnamefont{M.}~\bibnamefont{Arndt}},
  \bibinfo{author}{\bibfnamefont{M.~B.} \bibnamefont{Dahan}},
  \bibinfo{author}{\bibfnamefont{D.}~\bibnamefont{Gu\'ery-Odelin}},
  \bibinfo{author}{\bibfnamefont{M.~W.} \bibnamefont{Reynolds}},
  \bibnamefont{and} \bibinfo{author}{\bibfnamefont{J.}~\bibnamefont{Dalibard}},
  \bibinfo{journal}{Phys. Rev. Lett.} \textbf{\bibinfo{volume}{79}},
  \bibinfo{pages}{625} (\bibinfo{year}{1997}).

\bibitem[{\citenamefont{Burt et~al.}(1997)\citenamefont{Burt, Ghrist, Myatt,
  Holland, Cornell, and Wieman}}]{burt97}
\bibinfo{author}{\bibfnamefont{E.~A.} \bibnamefont{Burt}},
  \bibinfo{author}{\bibfnamefont{R.~W.} \bibnamefont{Ghrist}},
  \bibinfo{author}{\bibfnamefont{C.~J.} \bibnamefont{Myatt}},
  \bibinfo{author}{\bibfnamefont{M.~J.} \bibnamefont{Holland}},
  \bibinfo{author}{\bibfnamefont{E.~A.} \bibnamefont{Cornell}},
  \bibnamefont{and} \bibinfo{author}{\bibfnamefont{C.~E.}
  \bibnamefont{Wieman}}, \bibinfo{journal}{Phys. Rev. Lett.}
  \textbf{\bibinfo{volume}{79}}, \bibinfo{pages}{337} (\bibinfo{year}{1997}).

\bibitem[{\citenamefont{Moerdijk et~al.}(1996)\citenamefont{Moerdijk, Boesten,
  and Verhaar}}]{moerdijk96}
\bibinfo{author}{\bibfnamefont{A.~J.} \bibnamefont{Moerdijk}},
  \bibinfo{author}{\bibfnamefont{H.~M. J.~M.} \bibnamefont{Boesten}},
  \bibnamefont{and} \bibinfo{author}{\bibfnamefont{B.~J.}
  \bibnamefont{Verhaar}}, \bibinfo{journal}{Phys. Rev. A}
  \textbf{\bibinfo{volume}{53}}, \bibinfo{pages}{916} (\bibinfo{year}{1996}).

\bibitem[{\citenamefont{Weber et~al.}(2002)\citenamefont{Weber, Herbig, Mark,
  Naegerl, and Grimm}}]{weber02}
\bibinfo{author}{\bibfnamefont{T.}~\bibnamefont{Weber}},
  \bibinfo{author}{\bibfnamefont{J.}~\bibnamefont{Herbig}},
  \bibinfo{author}{\bibfnamefont{M.}~\bibnamefont{Mark}},
  \bibinfo{author}{\bibfnamefont{H.-C.} \bibnamefont{Naegerl}},
  \bibnamefont{and} \bibinfo{author}{\bibfnamefont{R.}~\bibnamefont{Grimm}},
  \bibinfo{journal}{Science} \textbf{\bibinfo{volume}{299}},
  \bibinfo{pages}{232} (\bibinfo{year}{2002}).

\bibitem[{\citenamefont{Katori et~al.}(1999)\citenamefont{Katori, Ido, Isoya,
  and Kuwata-Gonokami}}]{katori99}
\bibinfo{author}{\bibfnamefont{H.}~\bibnamefont{Katori}},
  \bibinfo{author}{\bibfnamefont{T.}~\bibnamefont{Ido}},
  \bibinfo{author}{\bibfnamefont{Y.}~\bibnamefont{Isoya}}, \bibnamefont{and}
  \bibinfo{author}{\bibfnamefont{M.}~\bibnamefont{Kuwata-Gonokami}},
  \bibinfo{journal}{Phys. Rev. Lett.} \textbf{\bibinfo{volume}{82}},
  \bibinfo{pages}{1116} (\bibinfo{year}{1999}).

\end{thebibliography}
\end{document}